\documentclass[a4paper,11pt,singlespace]{article}
\usepackage[utf8]{inputenc}

\usepackage{lineno}
\usepackage{subfig}
\usepackage{amsmath}
\usepackage{graphicx}
\usepackage[left=2.5cm,right=2.5cm]{geometry}
\usepackage[hidelinks]{hyperref}

\title{Signal and Power transmission over Fiber \\in the DUNE Far Detector\\\small Contribution to the 25th International Workshop on Neutrinos from Accelerators}
\author{Sabrina Sacerdoti - for the DUNE Collaboration\\ \textit{Laboratoire AstroParticule et Cosmologie - CNRS}}
\date{\vspace{-5ex}}

\begin{document}

\maketitle

\begin{abstract}
\begin{footnotesize}
  One of the far detectors of the Deep Underground Neutrino Experiment (DUNE) will be instrumented with a Vertical Drift Liquid Argon Time Projection Chamber (VD LArTPC). It will also be equipped with a Photo-Detection System (PDS), which determines the time of interaction of non-beam events, as well as contributing to the energy reconstruction. The characteristics of this new type of LArTPC required an innovative approach to the placement of the photo-detectors. In order to achieve a high light yield and uniform coverage of the detector, the PDS detectors in the VD LArTPC will be placed not only on the membrane walls of the cryostat, but also on the high voltage surface of the TPC's cathode. Such placement enhances the coverage of the PDS but presents an important technical challenge, since the powering and signal readout of the detectors must be done using non-conducting materials. To this end the Power-over-Fiber and Signal-over-Fiber transmission technologies were developed within the DUNE collaboration, opening a door to further enhance the coverage of future PD systems. This document will describe the technical solutions developed to this end, as well as presenting the final performance results from laboratory characterization of the devices and the prototype testing campaigns carried out at the CERN Neutrino Platform.
\end{footnotesize}
\end{abstract}

\section{The Vertical Drift Photo-Detection System}
\begin{figure}[htpb]
\centering
\subfloat[The Vertical Drift detector]{
\includegraphics[width=.5\textwidth]{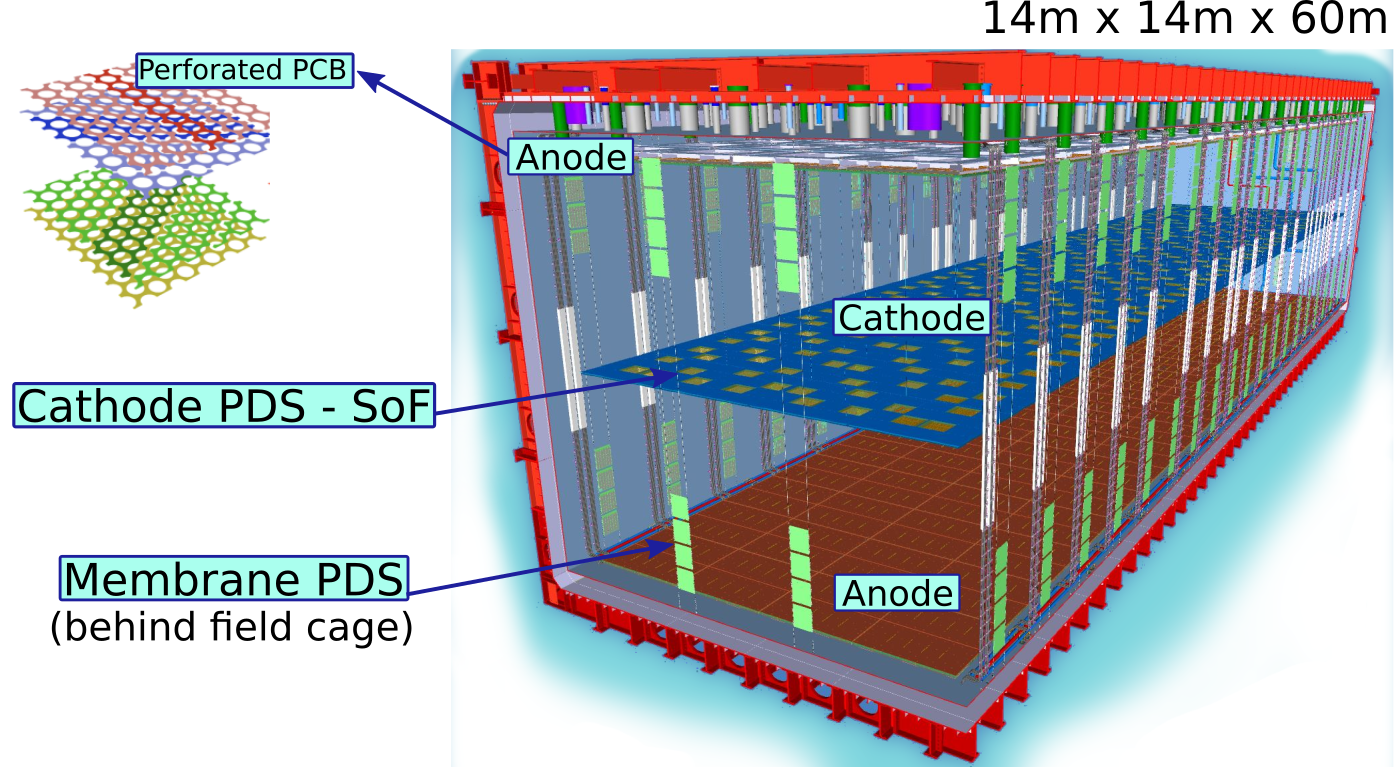}\label{fig:VDdetector}}
\subfloat[Simulated light yield~\cite{VD-TDR}.]{
\includegraphics[width=.4\textwidth]{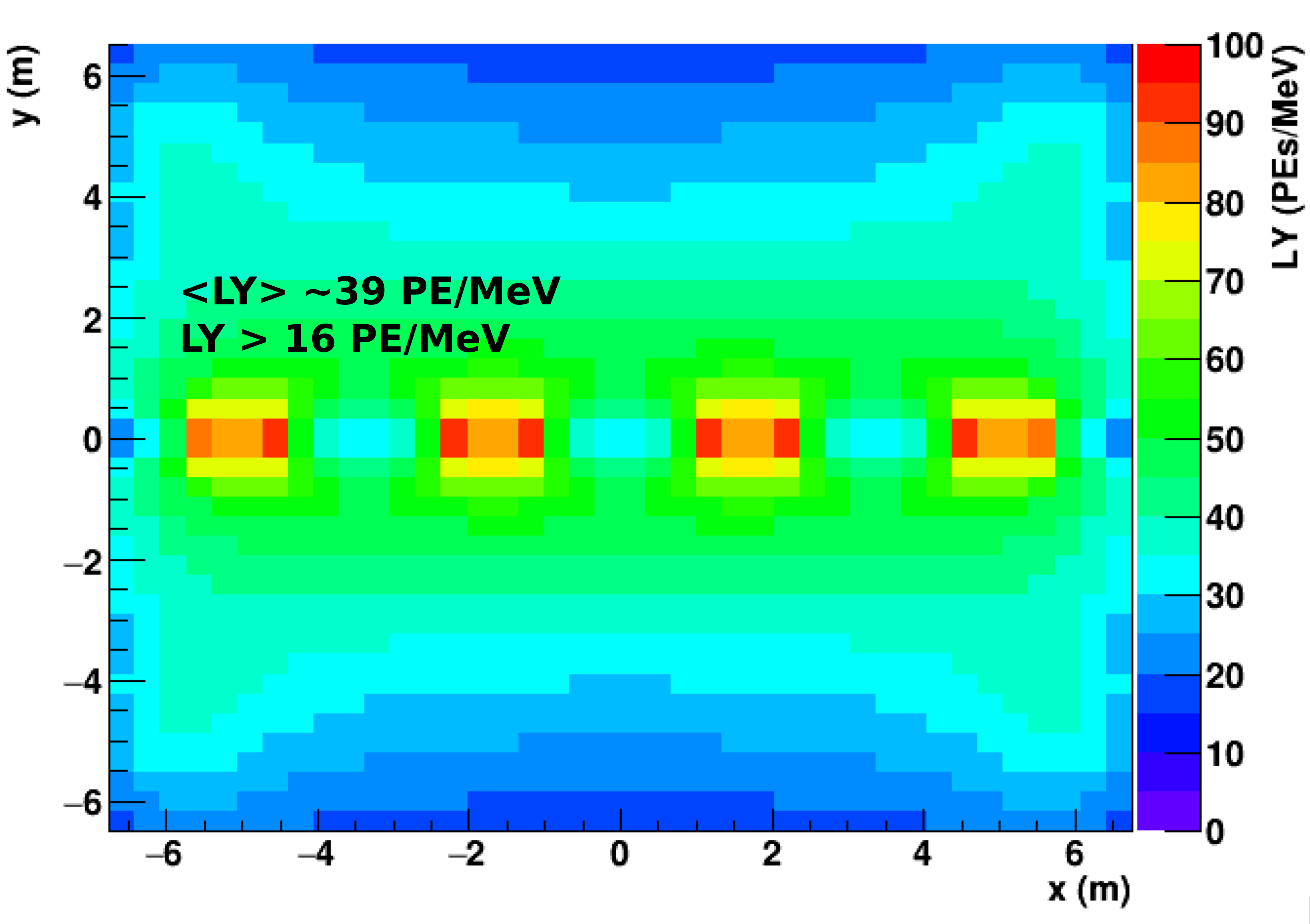}\label{fig:l-yield}}

\caption{Schematic of the Vertical Drift Far Detector, with the positions of the cathode and membrane PDS detectors highlighted. ~\ref{fig:l-yield} presents the light yield simulation results in a transversal cut of the detector.}
\label{fig:vd-schem}
\end{figure}

The Deep Underground Neutrino Experiment (DUNE) is a next-generation long-baseline neutrino experiment that aims to address some of the main open questions in particle physics. Its Far Detector will have a 1300~km baseline and will be located at the Sanford Underground Research Facility in South Dakota, 1.5 km deep underground. It will count with four modules, one of which will be a Vertical Drift LArTPC~\cite{VD-TDR}. In this type of TPC, the ionization electrons drift vertically in the 500~V/cm electric field generated by the cathode, which is suspended at mid-height of the detector and biased to -300~kV. The anodes are located 6~m away, at the top and bottom of the detector. A schematic drawing of the detector is shown in Fig.~\ref{fig:VDdetector}.

The detector is also instrumented with a Photo-Detection System (PDS), which serves to provide the timing information of the neutrino interactions. Half of the photo-detectors are placed on the walls of the cryostat, behind the field cage. In order to increase the light-yield and improve the light-detection uniformity, the second half is placed on the high-voltage surface of the cathode. Their effect can be appreciated in the simulation results shown in Fig.~\ref{fig:l-yield}.



The photo-detectors, called xArapuca~\cite{arapuca}, are structured around a large, $60 \times 60\, \text{cm}^2$ wavelength shifter plate, that guides the collected light towards the 160 Silicon Photo Multipliers (SiPM) positioned around it. The SiPMs are mounted in groups of 20 on flex boards, and grouped into two readout channels. The cathode detectors are double-sided, and see light from the volumes above and below them. However, they are agnostic to the direction and thus require matching to the membrane modules in order to determine it.

Placing photo-detectors on a high-voltage surface required the development of systems to both power and read-out the sensors using only non-conductive materials. Two technologies were thus developed within the DUNE PDS consortium. The first, Power-over-Fiber\cite{pof}, allows to power the readout electronics and bias the SiPMs. The Signal-over-Fiber, which is the main focus of this document, enables the transmission of the SiPM signals using optical fibers.

\section{Power-over-Fiber}
The light of a high power (1~W) infra-red (IR) laser is transmitted using multi-mode (MM) optical fibers from the outside of the cryostat towards the location of the detector. The electronics are then equipped with InGaAs optical power converters (OPC) that have been optimized for cold operation and can efficiently convert the power from the light into electrical power.

Although the SiPM peak sensitivity is far away from IR, the power of the PoF laser is such that light leaks from the fiber or the OPCs could highly affect the detector's operation by generating a high single photo-electron (SPE) count. This is why special care has been taken to select fibers with black tubing and metallic FC connectors, protected by meshes and additional black covers. Further information in ~\cite{pof}.

\section{Signal-over-Fiber}

\begin{figure}
\centering
  \includegraphics[width=.8\textwidth]{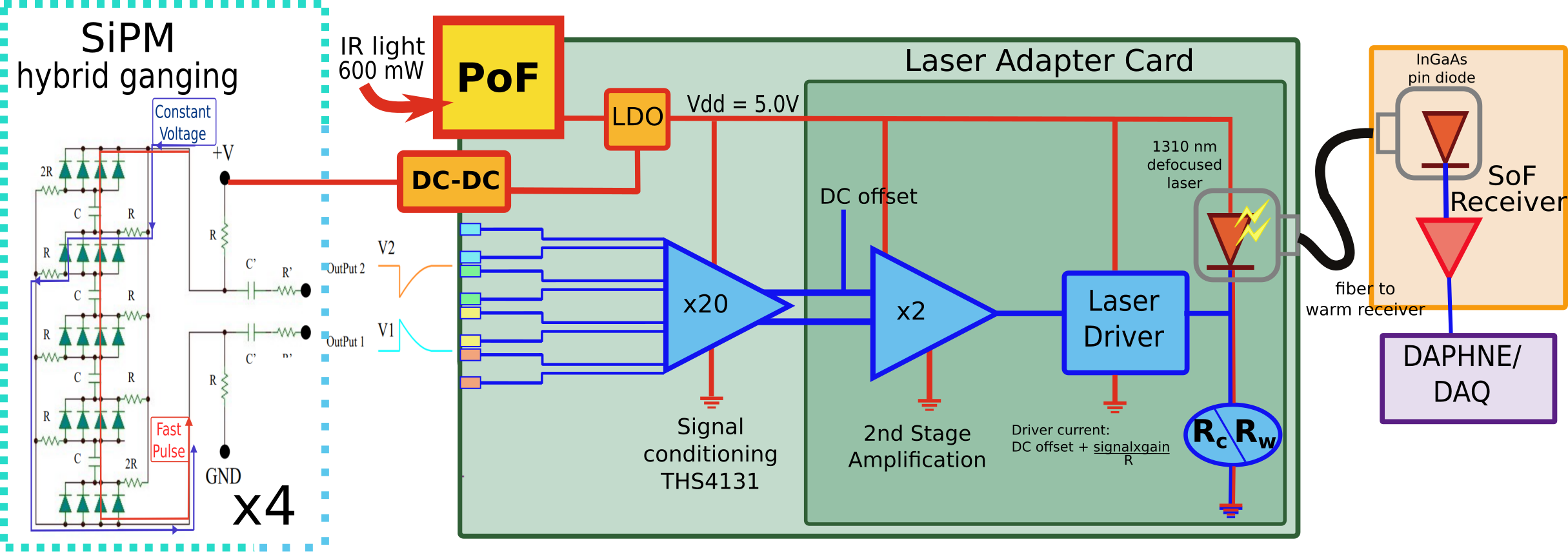}
  \caption{Schematic drawing of the SoF readout electronics, with the SiPM ganging on the left. The cryogenic circuit: the signal conditioning and laser driver. An optical fiber carries the signal from the IR lasers towards the warm optical receiver and digitization electronics on the right.}
  \label{fig:sof-circuit}
\end{figure}

The SoF electronics consist of a cryogenic analogue optical transmitter that converts the SiPM signals into light, that is guided using MM optical fibers towards the outside of the cryostat to the SoF warm analogue optical receiver. The signals are then converted back to electrical signals to be digitized by the DAPHNE digitization electronics. A schematic of the SoF electronics is shown in Fig.~\ref{fig:sof-circuit}.

The SiPMs in the xArapucas are interconnected in a \textit{hybrid ganging} scheme that allows to have the bias and differential readout on the same line. Four groups of 20 SiPMs are AC-coupled to each channel. As explained before, PoF is used to power the circuit, through a low drop-out voltage regulator that outputs a stable Vcc. It also powers an in-house designed DC-DC voltage step-up circuit that provides the bias voltage for the SiPMs.

The circuit contains two amplification stages, with a gain tuned to optimize signal-to-noise ratio (SNR) within the available voltage range; a compromise is necessary between SNR and dynamic range. A constant DC offset is generated to keep the laser above the lasing threshold. A last stage consisting of a laser driver converts the signals from voltage to current, driving the IR laser. The power output of laser diodes is linear to current, above a threshold; Interestingly, this threshold is around 10~mA at room temperature, and drops to 2-3~mA when in LAr.  A negative coefficient resistor in the laser driver allows to switch between warm and cold operation automatically.

\subsection{Prototyping}
The evaluation and validation of the performance of the full integral PDS is done at the CERN Neutrino Platform, in a $3\times 3 \times 1\, \text{m}^3$ cryostat. This facility permits the testing of test full-size PDS components in conditions similar to the real detector, and alongside the TPC. The cathode can be biased up to -30~kV. Measurements are done both with cosmic muons simultaneously detected by the TPC and the PDS, as well as with a calibration system equipped with a UV LED flasher. More than 15 data-taking runs have happened since November 2021.

\begin{figure}[htpb]
\centering
\subfloat[Small signal charge distribution]{
\includegraphics[width=.45\textwidth]{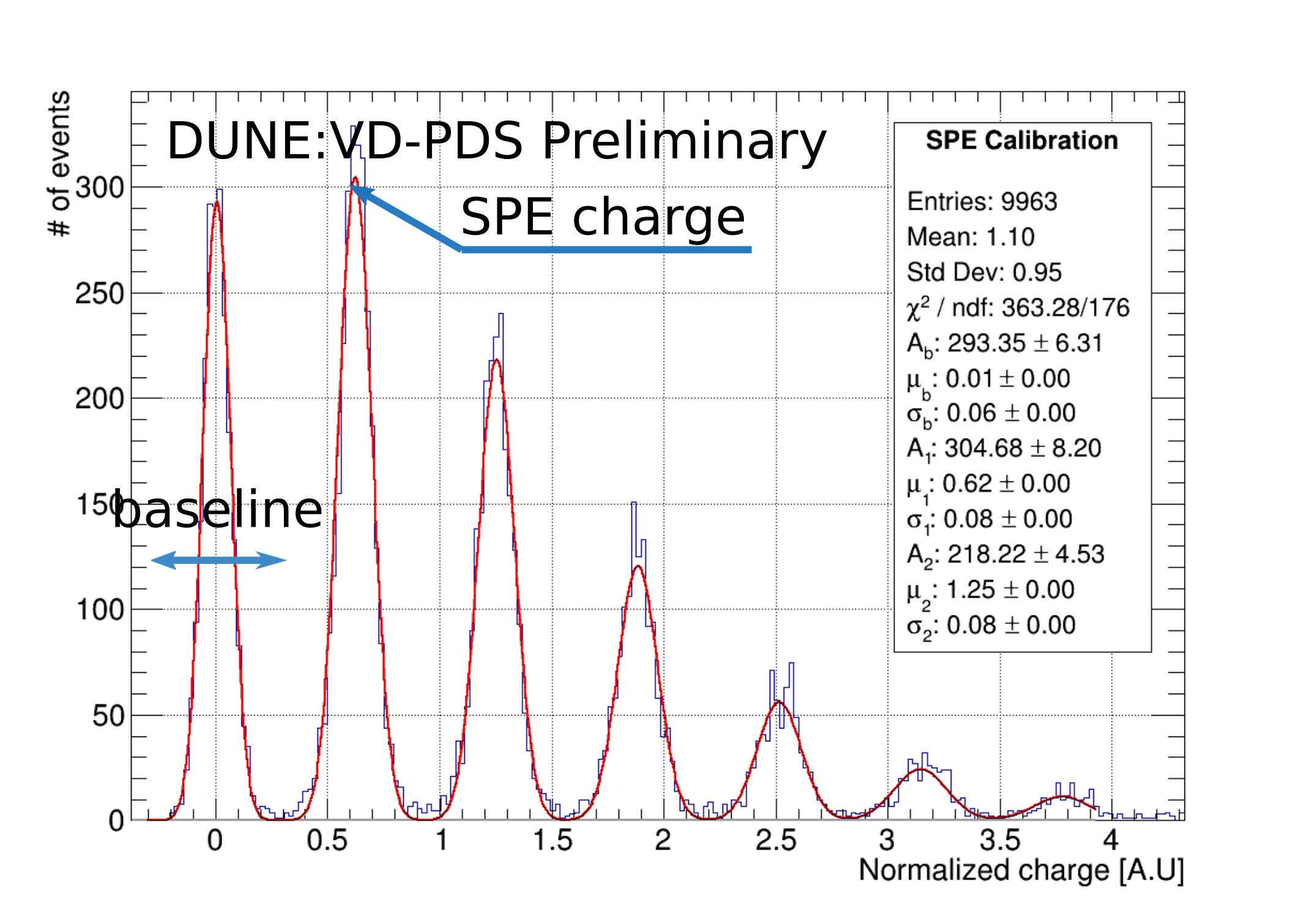}\label{fig:spe}}
\subfloat[Dynamic range]{
\includegraphics[width=.55\textwidth]{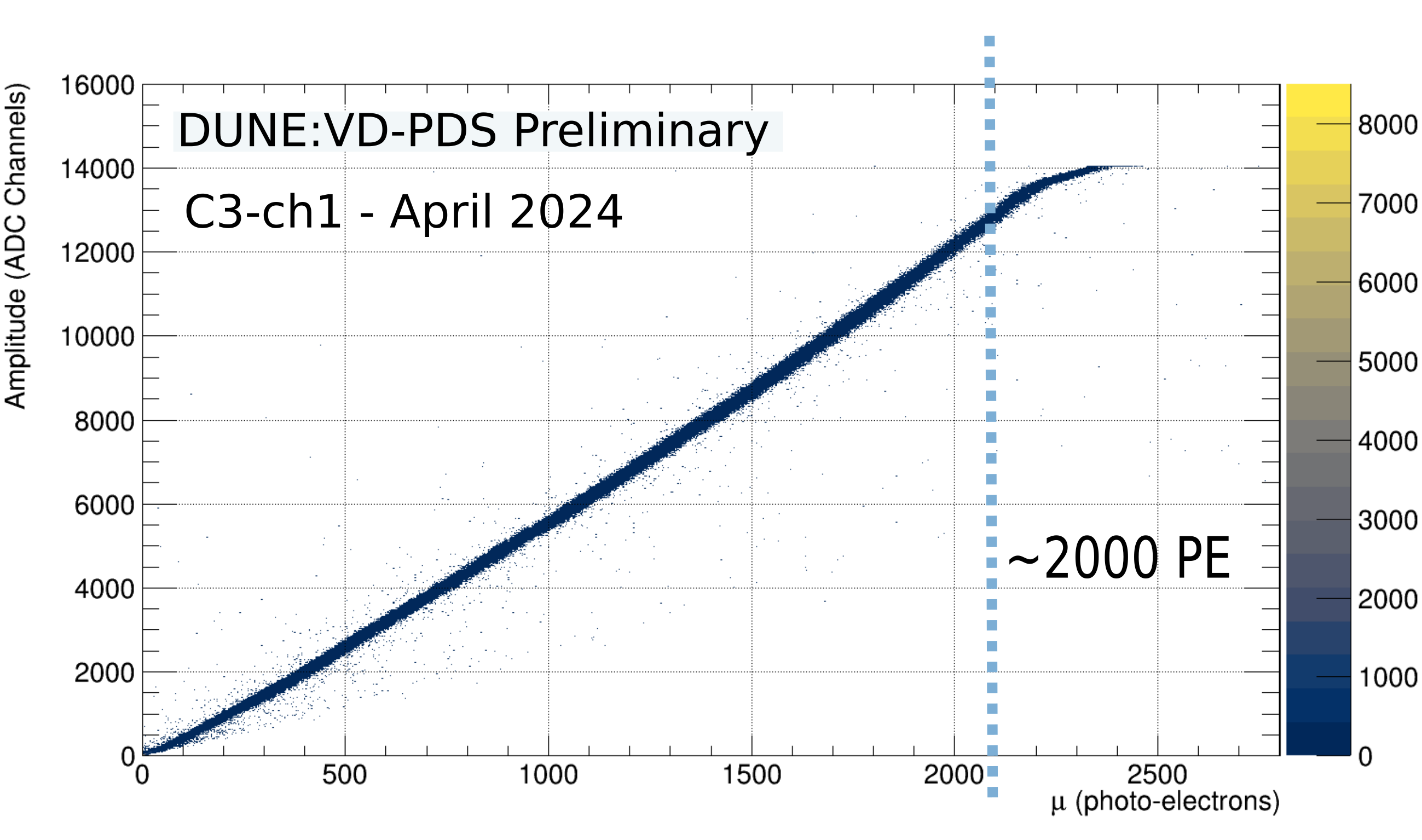}\label{fig:dynamic}}

\caption{Results reflecting the performance of the cold electronics in the prototyping run of April 2024.}
\label{fig:results}
\end{figure}

The two figures of merit by which the system is evaluated are the SNR, which evaluates how well the system can transmit an SPE signal, and the dynamic range, which shows what's the largest signal the electronics could transmit without saturating. Fig.~\ref{fig:results} shows examples of results from the last data taking in April 2024. In Fig.~\ref{fig:spe} the charge histogram of few-PE signals is shown, based on which the SNR is calculated. Fig.~\ref{fig:dynamic} shows the amplitude of the signal as a function of the estimated PE (based on the SPE charge). Four xArapucas (8 channels) were tested, all achieving SNR$>$6 and a dynamic range between 1600 and 2000 PE.

The warm SoF receiver was first tested in this run as well, and its integration to the DAPHNE electronics is on-going.

%

\section{Conclusions}

The geometry of the VD-LArTPC presented a challenge for the placement of the PDS sensors. Simulation studies showed that positioning sensors on the cathode would result in a highly improved light-yield distribution. In order to make this possible, two novel systems have been developed to power and read-out these detectors.

The SoF electronics have been tested in prototype runs at CERN in multiple occasions. In 2024, 8 channels were run simultaneously inside of the test cryostat, and 16 channels were tested for installation in ProtoDUNE. The overall performance shown was consistently meeting the DUNE FD2 requirements.

\end{document}